\documentclass[showpacs, pra, superscriptaddress, aps, twocolumn,10pt]{revtex4}
\usepackage{mathrsfs}
\usepackage{array}
\usepackage{amsmath}
\usepackage{graphicx}
\usepackage{amstext}
\usepackage{bm}
\usepackage{amsfonts}

\begin{document}
\title{Einstein-Podolsky-Rosen entanglement and asymmetric steering between distant macroscopic mechanical and magnonic systems 
}
\author{Huatang Tan}
\email{tht@mail.ccnu.edu.cn}
\affiliation{Department of Physics, Huazhong Normal University, Wuhan 430079, China}
\author{Jie Li}
\email{J.Li-17@tudelft.nl}
\affiliation{Zhejiang Province Key Laboratory of Quantum Technology and Device,
Department of Physics and State Key Laboratory of Modern Optical Instrumentation, Zhejiang University, Hangzhou 310027, China}
\affiliation{Kavli Institute of Nanoscience, Department of Quantum Nanoscience,
Delft University of Technology, 2628CJ Delft, The Netherlands}

\begin{abstract}
We propose a deterministic scheme for establishing hybrid Einstein-Podolsky-Rosen (EPR) entanglement channel between a macroscopic mechanical oscillator and a magnon mode in a distant yttrium-iron-garnet (YIG) sphere across about ten gigahertz of frequency difference. The system consists of a driven electromechanical cavity which is unidirectionally coupled to a distant electromagnonical cavity inside which a YIG sphere is placed. We find that far beyond the sideband-resolved regime in the electromechanical subsystem, stationary phonon-magnon EPR entanglement can be achieved. 
This is realized by utilizing the output field of the electromechanical cavity being an intermediary which distributes the electromechanical entanglement to the magnons, thus establishing a remote phonon-magnon entanglement. The EPR entanglement is strong enough such that phonon-magnon quantum steering can be attainable 
in an asymmetric manner. This long-distance macroscopic hybrid EPR entanglement and steering enable potential applications not only in fundamental tests of quantum mechanics at the macro scale, but also in quantum networking and one-sided device-independent quantum cryptography based on magnonics and electromechanics.
\end{abstract}
\maketitle

\emph{\textbf{Introduction.}}--Long-distance entanglement has attracted extensive attention owing to its potential applications to the fundamental test of quantum mechanics \cite{gis}, quantum networking \cite{le1,le2}, and quantum-enhanced metrology \cite{plzk1}. In such quantum tasks, hybrid quantum systems, composed of distinct physical components with complementary functionalities, possess multitasking capabilities and thus may be better suited than others for specific tasks \cite{kuk}. It has been proved that light mediation is an effective approach to achieving the entanglement between two remote systems that never interact directly \cite{sid, jod, plzk2}. For example, hybrid EPR entanglement between distant macroscopic mechanical and atomic systems has been realized very recently via unidirectional light coupling \cite{plzk2}. In addition, strong coherent coupling between a mechanical membrane and atomic spins has been demonstrated by two cascade light-mediated coupling processes \cite{ham}.

On the other hand, the realization of nonclassical effects of macroscopic objects is an ongoing effort in quantum science \cite{gisin}, due to their great use for, e.g., testing the validity of quantum mechanics \cite{kim, milburn} and probing decoherence theories \cite{zk,Bassi} at large mass scales. In cavity optomechanics and electromechanics, which involve the hybrid coupling of massive mechanical resonators to electromagnetic field \cite{pt,com}, recent experiments have succeeded in preparing a variety of quantum states of macroscopic mechanical oscillators \cite{squ1, squ2, entanglement1, entanglement2, omen1, omen2, markus, spn, omb}, including quantum squeezing and entanglement of mechanical oscillators \cite{squ1, squ2, entanglement1, entanglement2}, nonclassical correlations between photons and phonons \cite{omen1,markus}, single-phonon Fock states \cite{spn}, and optomechanical Bell nonlocality \cite{omb}, etc. These macroscopic quantum states can also be useful in the processing and communication of quantum information \cite{rip} and ultrahigh precision measurement beyond the standard quantum limit \cite{cav}. In addition, hybrid interfaces of mechanical oscillators with other systems such as atomic ensembles \cite{plzk2, ham}, nitrogen vacancy centers \cite{arc}, and superconducting devices \cite{cleland} have already been realized \cite{asp}.

Apart from optomechanical and electromechanical systems, hybrid systems based on magnons in macroscopic magnetic materials have increasingly become a new and promising platform for studying macroscopic quantum effects, attributed to magnon's great frequency tunability, very low damping loss, and excellent coupling capability to microwave or optical photons, phonons and qubits \cite{mag}.
Experiments have realized strong cavity-magnon coupling \cite{mpo1,mpo2,mpo3, mpo4} in YIG spheres and other related interesting phenomena, such as  magnon gradient memory \cite{tang}, exceptional point \cite{expo}, bistability \cite{bis} and nonreciprocity and unidirectional invisibility in cavity magnonics \cite{hu}. Moreover, entanglement-based single-shot detection of a single magnon with a superconducting qubit has been realized \cite{esd}. Recent schemes for achieving quantum phenomena of magnons, including squeezing and entanglement, quantum steering, magnon blockade, and magnon-mediated microwave entanglement, have already been proposed \cite{li1, li2, scully, tan2, ying, fuli, ficek, li3}.

Here we consider a microwave-mediated phonon-magnon interface and focus on how to deterministically establish hybrid EPR entanglement channel between a macroscopic mechanical oscillator and a {\it distant} YIG sphere across about ten gigahertz of frequency difference. The system consists of two unidirectionally-coupled electromechanical and electromagnonical cavities inside which a mechanical oscillator and a YIG sphere are placed, respectively.  We find that far beyond the electromechanical sideband-resolved regime, strong stationary phonon-magnon EPR entanglement and steering can be achieved, as a result of electromechanical output photon-phonon entanglement distributed via the unidirectional cavity coupling. Further, the one-way steering from phonons to magnons is established and adjustable over a wide range of feasible parameters. The entanglement and steering are robust against the frequency dismatch between the two cavities, unidirectional cavity-coupling loss, and environmental temperature.

\begin{figure}[t]
\centerline{\scalebox{0.24}{\includegraphics{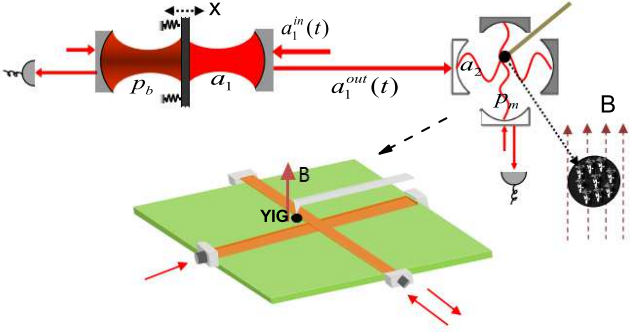}}}
 \caption{Schematic diagram. A driven electromechanical cavity ($a_1$) is unidirectionally coupled to an electromagnonical cavity ($a_2$) where a YIG sphere in a uniform magnetic field is placed. The probing cavities $p_{b}$ and $p_{m}$ are used to read out and detect the phonon-magnon entangled state. The two-cavity configuration ($a_2,~p_m$) of the microwave optomagnonic system can be a cross-shape cavity \cite{Rao}, and the YIG sphere is glued
on the end of a cantilever.}
 \label{sys}
\end{figure}

\emph{\textbf{Model.}}--We consider a driven electromechanical cavity that is unidirectionally coupled to an electromagnonical cavity, as shown in Fig.\ref{sys}. For the electromechanical cavity, the cavity resonance is modulated by the motion of the mechanical oscillator, giving rise to the electromechanical coupling. Inside the electromagnonical cavity, a ferrimagnetic YIG sphere with a diameter about hundreds of micrometers is placed. The YIG sphere is also put in a uniform bias magnetic field, giving rise to magnetostatic modes of spin waves in the sphere. The cavities $p_{b,m}$, with their output fields sent to measurement apparatuses, are used to read out the phonon and magnon states, respectively, and the two cavities ($a_2,~p_m$) can be a cross-shape cavity coupled to the YIG sphere glued on the end of a cantilever \cite{Rao}. 
The magnons, which characterize quanta of the uniform magnetostatic mode (i.e., Kittel mode \cite{kitt}) in the YIG sphere, are coupled to the cavity mode via magnetic dipole interaction. In the rotating frame with respect to the frequency $\omega_d$ of the drive with amplitude $\mathcal E_d$, the system's Hamiltonian $\hat H=\hat H_{ab}+\hat H_{am}$ reads ($\hbar=1$)
\begin{subequations}
\begin{align}
\hat H_{ab}&=\delta_1 \hat a_1^\dag \hat a_1{+} \,\omega_b \hat b^\dag\hat b \,{+} \tilde{g}_{ab}\hat a_1^\dag \hat a_1(\hat b {+} \hat b^\dag) {-}i (\mathcal E_d^*\hat a_1{-} \mathcal E_d\hat a_1^\dag),\\
\hat H_{am}&=\Delta_2 \hat a_2^\dag \hat a_2 {+} \Delta_m\hat m^\dag \hat m \,{+} g_{am}(\hat a_2^\dag\hat m+\hat a_2\hat m^\dag),
\end{align}
\label{e1}
\end{subequations}
where the bosonic annihilation operators $\hat a_j$, $\hat m$ and $\hat b$ denote the cavity, magnon, and mechanical modes with frequencies $\omega_j$, $\omega_m$ and $\omega_b$, respectively. The detunings $\delta_1=\omega_1-\omega_d$, $\Delta_2=\omega_2-\omega_d$ and $\Delta_{m}=\omega_{m}-\omega_d$. The magnon frequency $\omega_m=\beta H_B$, where $\beta$ is the gyromagnetic ratio and $H_B$ is the strength of the uniform bias magnetic field. $\tilde{g}_{ab}$ represents the single-photon electromechanical coupling, and $g_{am}$ denotes the magnetic-dipole coupling between the cavity and magnons, $g_{am} \propto \sqrt{N}$ with $N$ being the number of spins. In the recent experiment \cite{mpo2}, a strong coupling, $g_{am}\sim 47$~MHz, much larger than the cavity and magnon linewidths about $2.7$~MHz and $1.1$~MHz, has been achieved, whereas the electromechanical coupling $\tilde g_{ab}$ is typically weak, but it can be enhanced by using a strong cavity drive. Ultrastrong electromechanical coupling in the linear regime has been reported \cite{uts}. Under a strong drive, Eq.(\ref{e1}) can be linearized by replacing the operators by $\hat o\rightarrow \langle \hat o\rangle_{\rm ss}+\hat o~(o=a_j, b, m)$, where $\langle \hat o\rangle_{\rm ss}$ denotes the steady-state amplitudes of the modes, leading to the linearized Hamiltonian
\begin{subequations}
\begin{align}
\hat H_{ab}^{\rm lin}&=\Delta_1\hat a_1^\dag \hat a_1+\omega_b \hat b^\dag\hat b+g_{ab}(\hat a_1+\hat a_1^\dag)(\hat b+\hat b^\dag),\label{21}\\
\hat H_{am}^{\rm lin}&=\Delta_2\hat a_2^\dag \hat a_2+\Delta_m\hat m^\dag \hat m+g_{am}(\hat a_2^\dag\hat m+\hat a_2\hat m^\dag ),
\end{align}
\label{e2}
\end{subequations}
where $\Delta_1=\delta_1+2\tilde{g}_{ab}\text{Re}[\langle \hat b\rangle_{ss}]$ and  $g_{ab}= \tilde{g}_{ab}\langle \hat a_1\rangle_{ss}$, with
\begin{align}
\langle\hat a_1\rangle_{ss}=\frac{2\mathcal E_d}{\kappa_1+2i\Delta_1},~~~~\langle \hat b\rangle_{ss}=-\frac{2i\tilde{g}_{ab}|\langle\hat a_1\rangle_{ss}|^2}{\gamma_b+2i\omega_b},
\end{align}
and $\kappa_1$ and $\gamma_b$ are the damping rates of the cavity $\hat a_1$ and mechanical mode, respectively.

The unidirectional coupling between the two cavity fields can be described as follows: the output field $\hat a_1^{\rm out}(t)$ is used as the input field $\hat a_2^{\rm in}(t)$ to drive the cavity $\hat a_2$ but not vice versa, i.e.,
\begin{align}
\hat a_2^{\rm in}(t)=\sqrt{\eta}\hat a_1^{\rm out}(t)+\sqrt{1-\eta}\hat a_\eta^{\rm in}(t),
\end{align}
where the transmission loss is taken into account, with the coupling efficiency $\eta$, and the output field $\hat a_1^{\rm out}(t)=\sqrt{\kappa_1}\hat a_1(t)+\hat a_1^{\rm in}(t)$. The thermal noise operators $\hat a_l^{\rm in}(t)~(l=1,\eta)$ satisfy the nonzero correlations $\langle \hat{a}_l^{\rm in}(t) \hat{a}_{l'}^{\rm in\dag}(t')\rangle=(\bar{n}_l^{\rm th} +1)\delta_{ll'}\delta (t-t')$ and $\langle  \hat{a}_l^{\rm in\dag}(t)\hat{a}_{l'}^{\rm in}(t')\rangle=\bar{n}_l^{\rm th}\delta_{ll'}\delta (t-t')$, where $\bar{n}_1^{ \rm th}=(e^{\frac{\hbar\omega_1}{k_BT_1}}-1)^{-1}$ \Big[$\bar{n}_{\eta}^{ \rm th}=(e^{\frac{\hbar\omega_2}{k_BT_2}}-1)^{-1}$\Big] is the equilibrium mean thermal photon number at environmental temperature $T_1$ ($T_2$) and $k_B$ the Boltzmann constant.
By using Eqs.(\ref{e2}) and including the dissipations and input noises of the system, the equations of motion are derived as
\begin{subequations}
\begin{align}
&\frac{d}{dt}\hat a_1=-(\frac{\kappa_1}{2}+i\Delta_1)\hat a_1-ig_{ab}(\hat b+\hat b^\dag)-\sqrt{\kappa_1}\hat a_1^{\rm in}(t),\\
&\frac{d}{dt}\hat a_2=-(\frac{\kappa_2}{2}+i\Delta_2)\hat a_1-ig_{am}\hat m-\sqrt{\eta\kappa_1\kappa_2}\hat a_1\nonumber\\
&~~~~~~~~~~-\sqrt{\eta\kappa_2}\hat a_1^{\rm in}(t)-\sqrt{(1-\eta)\kappa_2}\hat {a}_{\eta}^{\rm in}(t),\\
&\frac{d}{dt}\hat b=-(\frac{\gamma_b}{2}+i\omega_b)\hat a_1-ig_{ab}(\hat a_1+\hat a_1^\dag)-\sqrt{\gamma_b}\hat b_{\rm in}(t),\\
&\frac{d}{dt}\hat m=-(\frac{\gamma_m}{2}+i\Delta_m)\hat m-ig_{am}\hat a_2-\sqrt{\gamma_m}\hat m_{\rm in}(t),
\end{align}
\label{e4}
\end{subequations}
where $\kappa_2$ and $\gamma_m$ are the damping rates of the cavity $\hat a_2$ and magnon mode, respectively. The noise operators $\hat b_{\rm in}(t)$ and $\hat m_{\rm in}(t)$ are independent and satisfy the same correlations as $\hat a_{1,\eta}^{\rm in}(t)$, with the mean thermal excitation numbers $\bar n_{b,m}^{\rm th}$ at temperature $T_{b,m}$.

When starting from Gaussian states, the system governed by the linearized Eq.(\ref{e4}) evolves still in Gaussian, whose state is completely determined by the covariance matrix $\sigma_{jj'}=\langle \mu_j\mu_{j'}+\mu_{j'}\mu_j\rangle/2-\langle \mu_j\rangle\langle \mu_{j'}\rangle$, where $\mu=(\hat x_1, \hat p_1, \hat x_2, \hat p_2,\hat x_b,\hat p_b,\hat x_m,\hat p_m)$ with the quadrature operators defined by $\hat x=(\hat o+\hat o^\dag)/\sqrt{2}$ and $\hat p=-i(\hat o-\hat o^\dag)/\sqrt{2}$ ($o=a_1,a_2,b,m$). The covariance matrix $\hat \sigma$ satisfies
\begin{align}
\dot{\sigma}&=A\sigma+\sigma A^T+D,
\label{me3}
\end{align}
where the drift matrix
\begin{align}
A=&\left(
  \begin{array}{cccc}
      A_1 & 0 & A_{ab} &0\\
     A_{12} &  A_2& 0& A_{am} \\
    A_{ab} & 0 &  A_b& 0  \\
    0 & A_{am} & 0 & A_m \\
  \end{array}
\right).
\end{align}
Here $A_{x=\{1,2\}}=-\begin{pmatrix}\begin{smallmatrix}\kappa_x & -2\Delta_x \\2\Delta_x & \kappa_x\end{smallmatrix}\end{pmatrix}/2$, $A_{y=\{b,m\}}=-\begin{pmatrix}\begin{smallmatrix}\gamma_y & -2\Delta_y \\2\Delta_y & \gamma_y\end{smallmatrix}\end{pmatrix}/2$ with  $\Delta_{b} \equiv \omega_b$,
$A_{ab}=-\begin{pmatrix}\begin{smallmatrix}0 & 0 \\0 & 2G_{ab}\end{smallmatrix}\end{pmatrix}$,
$A_{am}=\begin{pmatrix}\begin{smallmatrix}0 &g_{\rm am} \\-g_{\rm am} & 0\end{smallmatrix}\end{pmatrix}$, and $A_{12}=-\sqrt{\eta\kappa_1\kappa_2}I$.
The diffusion matrix
\begin{align}
D=\begin{pmatrix}\begin{smallmatrix}D_1 &D_{12} \\D_{12} & D_2\end{smallmatrix}\end{pmatrix}
\oplus \begin{pmatrix}\begin{smallmatrix}D_b &0 \\0 & D_m\end{smallmatrix}\end{pmatrix},
\end{align}
with $D_1=\kappa_1(\bar n_{1}^{\rm th}+1/2)I$, $D_2=\kappa_2\big[\eta(\bar n_{1}^{\rm th}+1/2)+(1-\eta)(\bar n_{\eta}^{\rm th}+1/2)\big]I$, $D_{12}=\sqrt{\eta\kappa_1\kappa_2}I/2$, and $D_s=\gamma_s(\bar n_s^{\rm th}+1/2)I~(s=b,m)$.

We are interested in the quantum correlations in the steady states which can be solved by setting the left-hand side of Eq.(\ref{me3}) to be zero. Note that the stability of the present master-slave cascade system is merely determined by the stability of the electromechanical subsystem, since the master subsystem is not influenced by the salve subsystem and the latter only involves linear mixing of the cavity and magnon modes.  The stability is therefore guaranteed when all the eigenvalues of the drift matrix of the electromechanical subsystem $\tilde A_{ab}\equiv\begin{pmatrix}\begin{smallmatrix}A_1 & A_{ab} \\A_{ab} & A_b\end{smallmatrix}\end{pmatrix}$ have negative real parts.

\begin{figure}[t]
\centerline{\scalebox{0.32}{\includegraphics{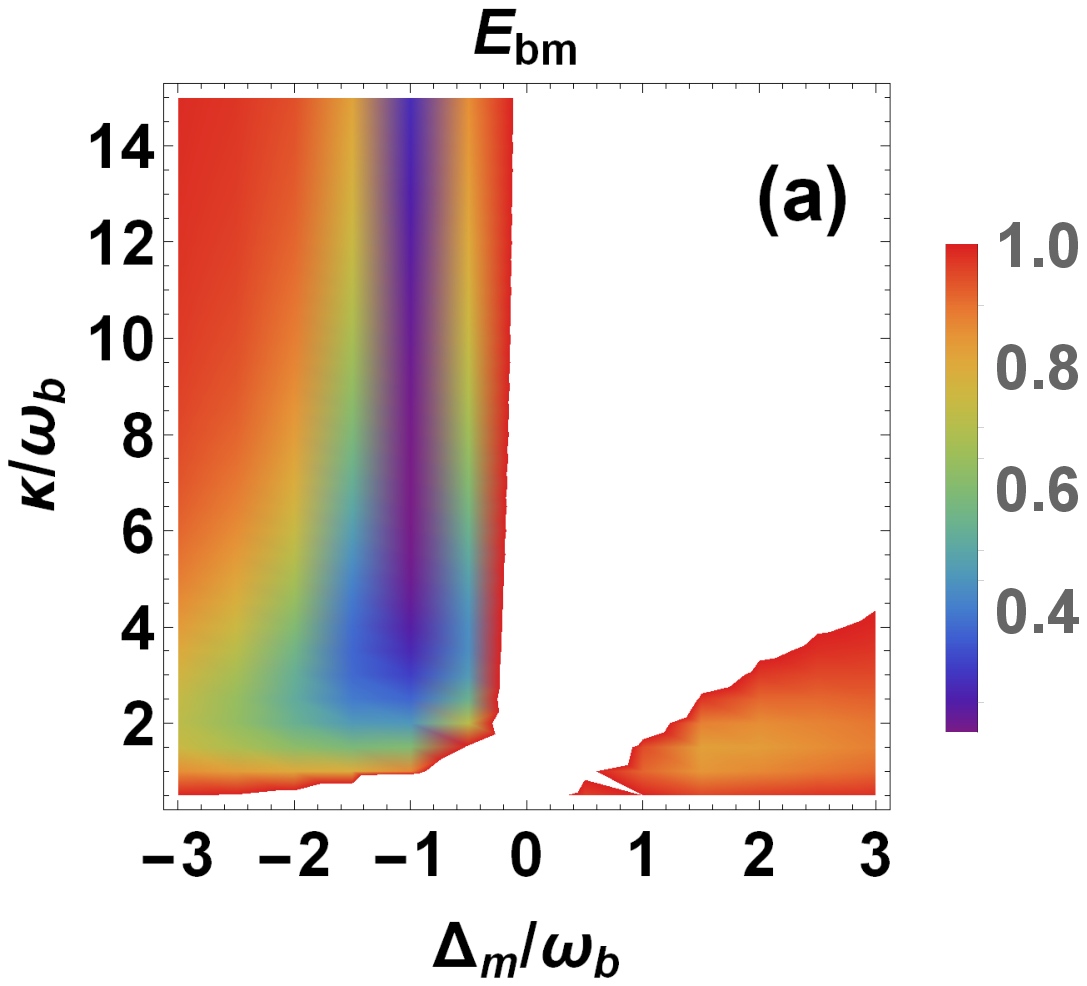}~~~~~~~~~~~~~\includegraphics{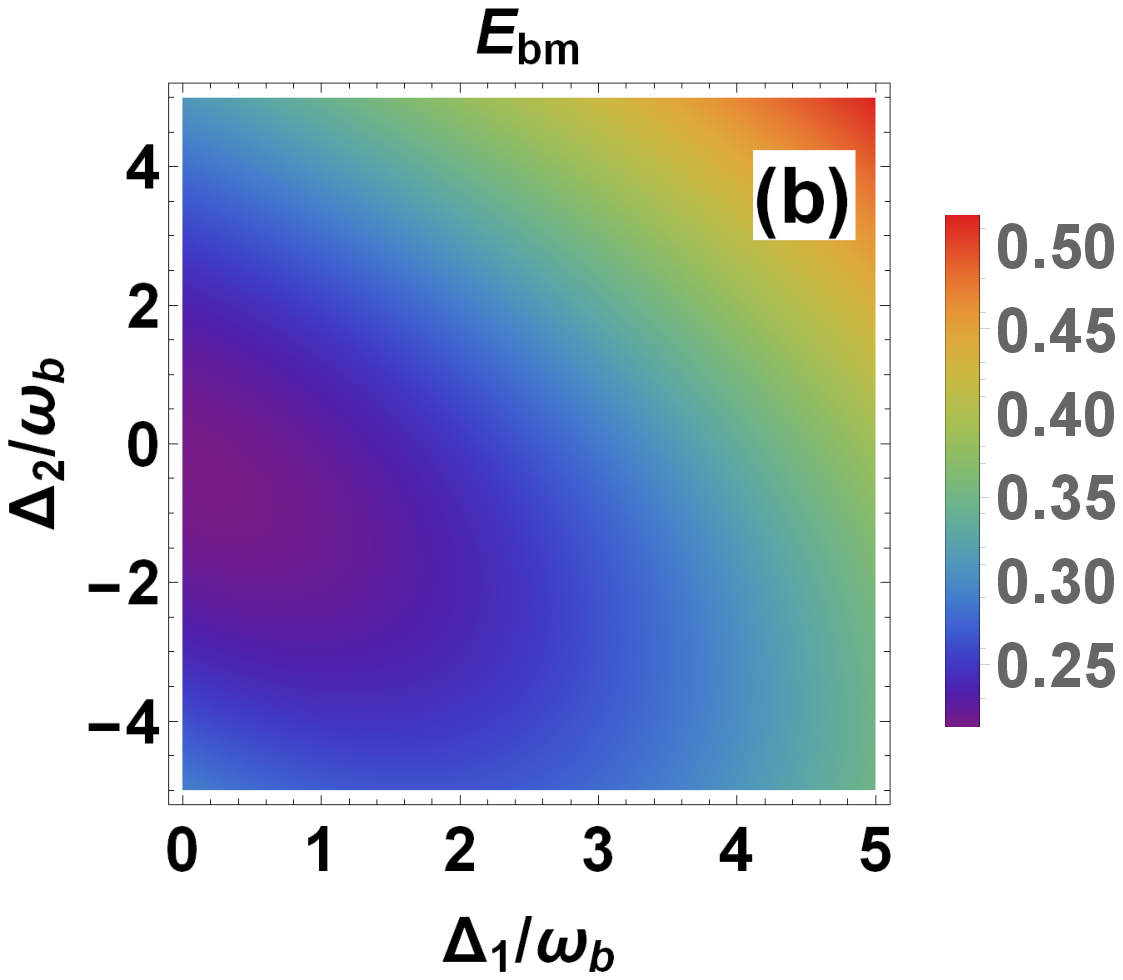}}}
\vspace{0.1cm}
\centerline{\scalebox{0.32}{\includegraphics{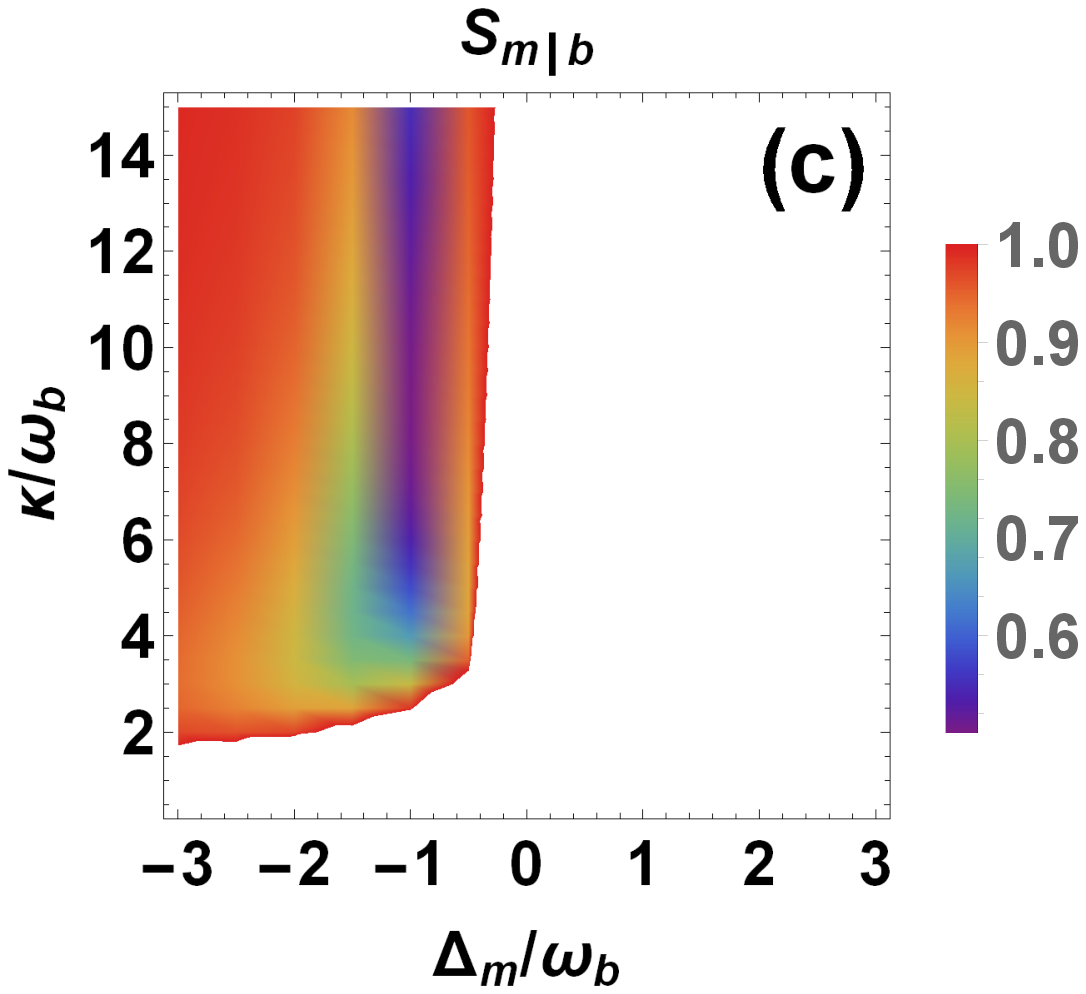}~~~~~~~~~~~~~\includegraphics{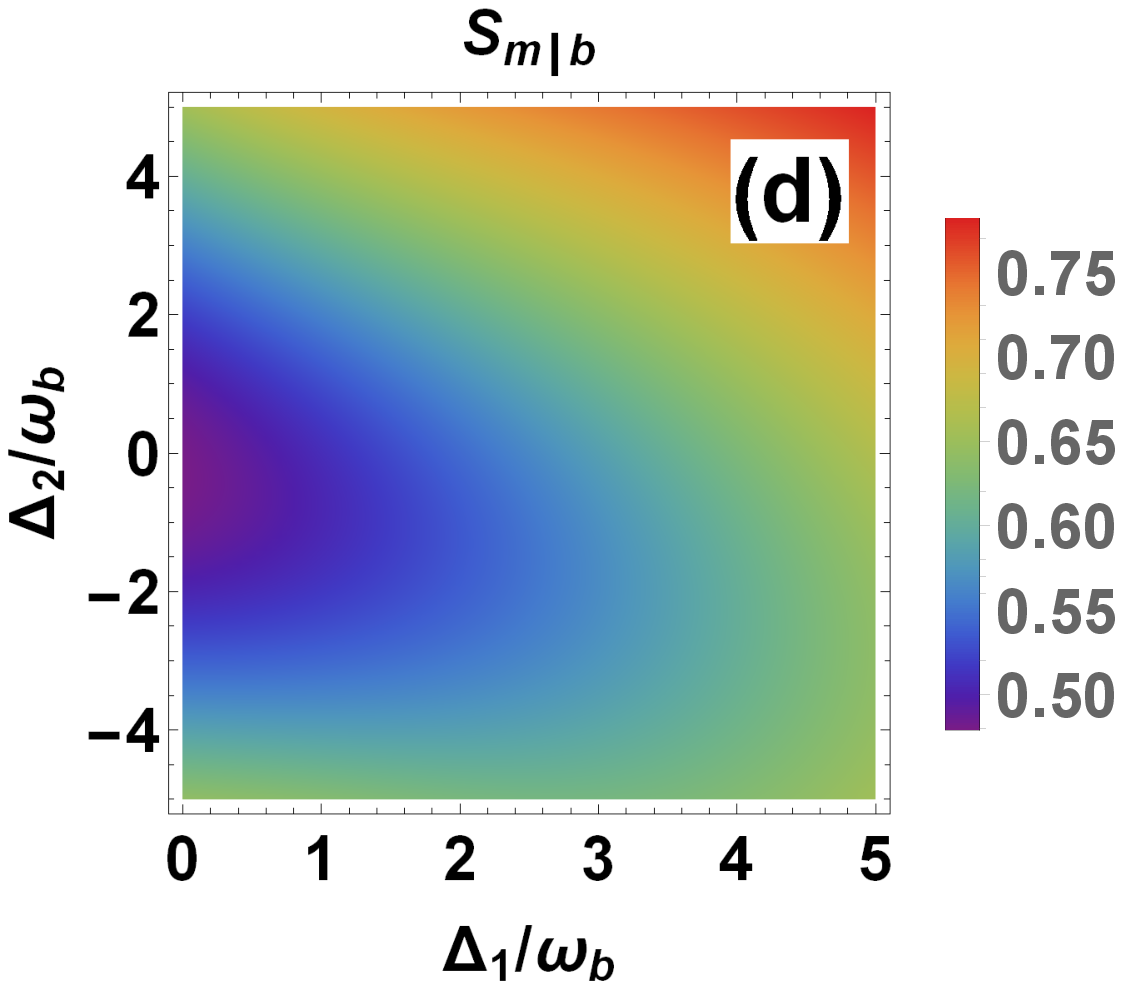}}}
 \caption{Density plots of the mechanical-magnon entanglement $E_{bm}$ and steering $S_{m|b}$ \emph{versus} the cavity dissipation rates $\kappa=\kappa_{1,2}$, the drive-magnon detuning $\Delta_m$, and the drive-cavity detuning $\Delta_{1,2}$. In (a) and (c), $\Delta_1=-\Delta_2=\omega_b$, and $g_{ab}=0.5g_{am}=0.5\omega_b$; in (b) and (d), $\kappa=10\omega_b$, $\Delta_m=-\omega_b$, and  $g_{ab}=0.5g_{am}=0.5\omega_b$.
 We take $\eta$=1 and the other parameters are provided in the text.
 In the plots, the steering $S_{b|m}$ is absent and not plotted, and the blank areas mean the absence of the entanglement and steering (similarly hereinafter).}
 \label{f2}
\end{figure}

\begin{figure*}[t]
\centerline{\scalebox{0.32}{\includegraphics{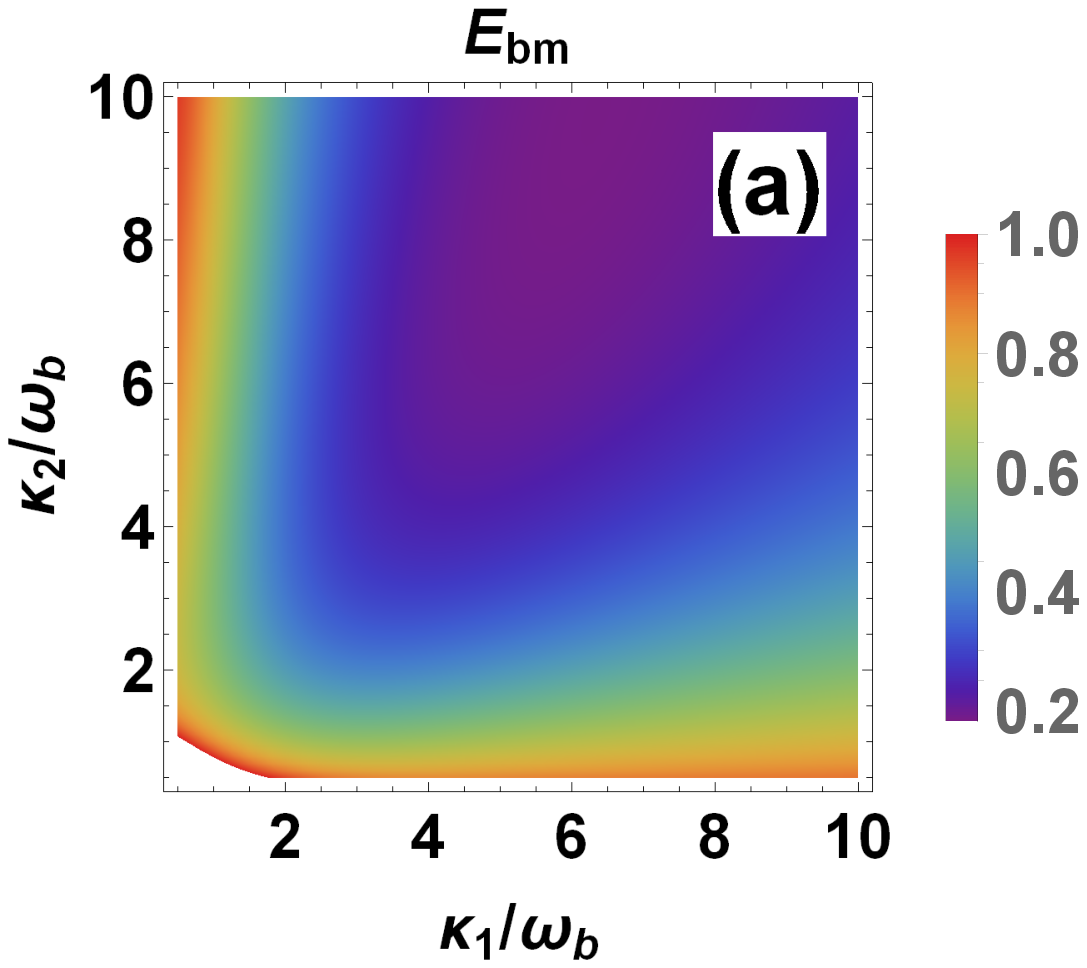}\hspace{1.5cm}\includegraphics{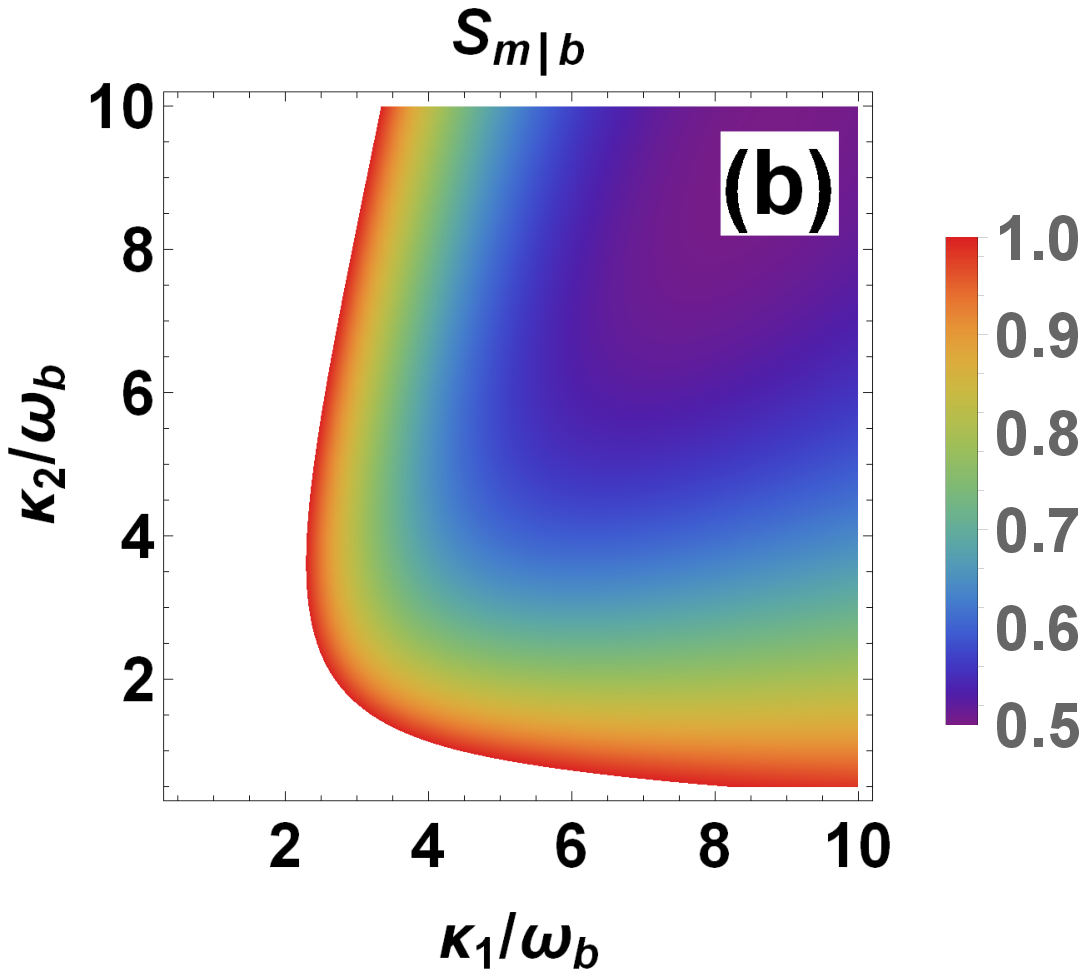}
\hspace{1.5cm}\includegraphics{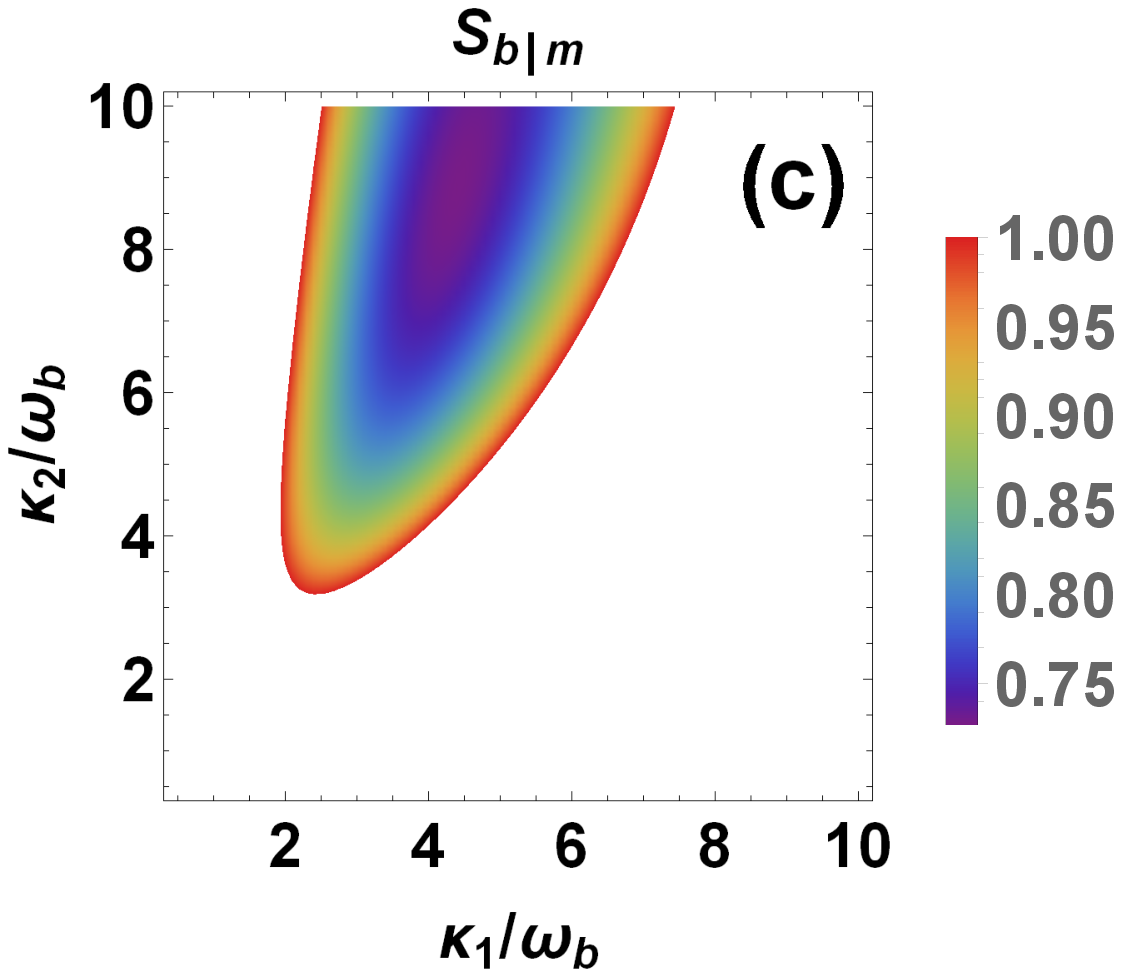}}}
\vspace{0.2cm}
\centerline{\scalebox{0.32}{\includegraphics{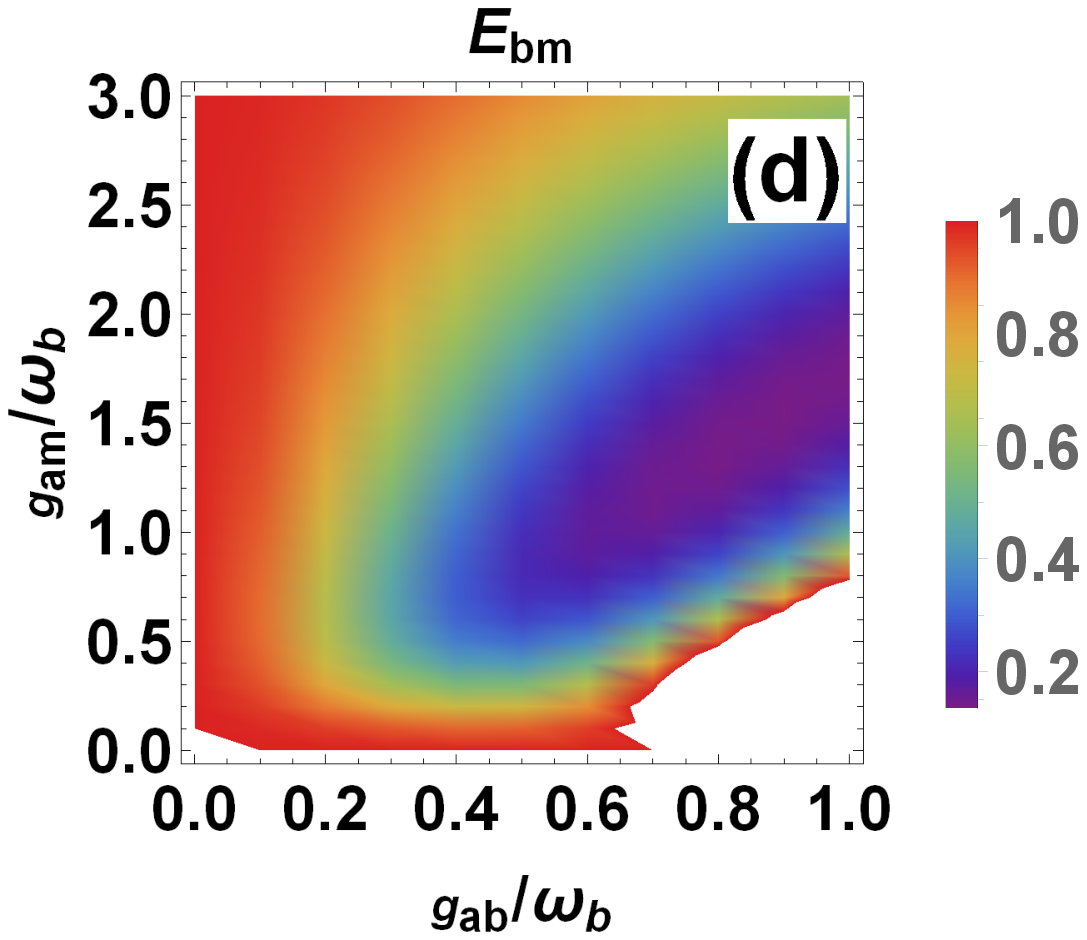}\hspace{1.5cm}\includegraphics{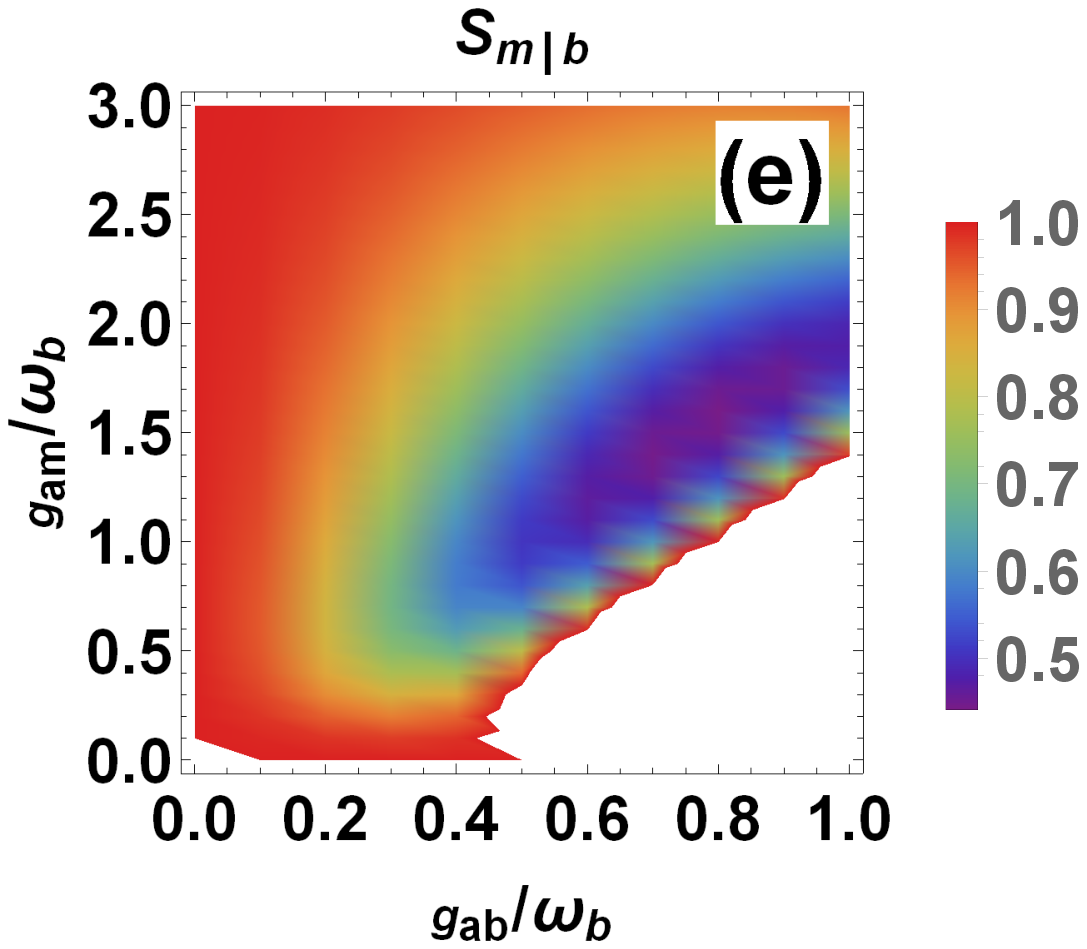}
\hspace{1.5cm}\includegraphics{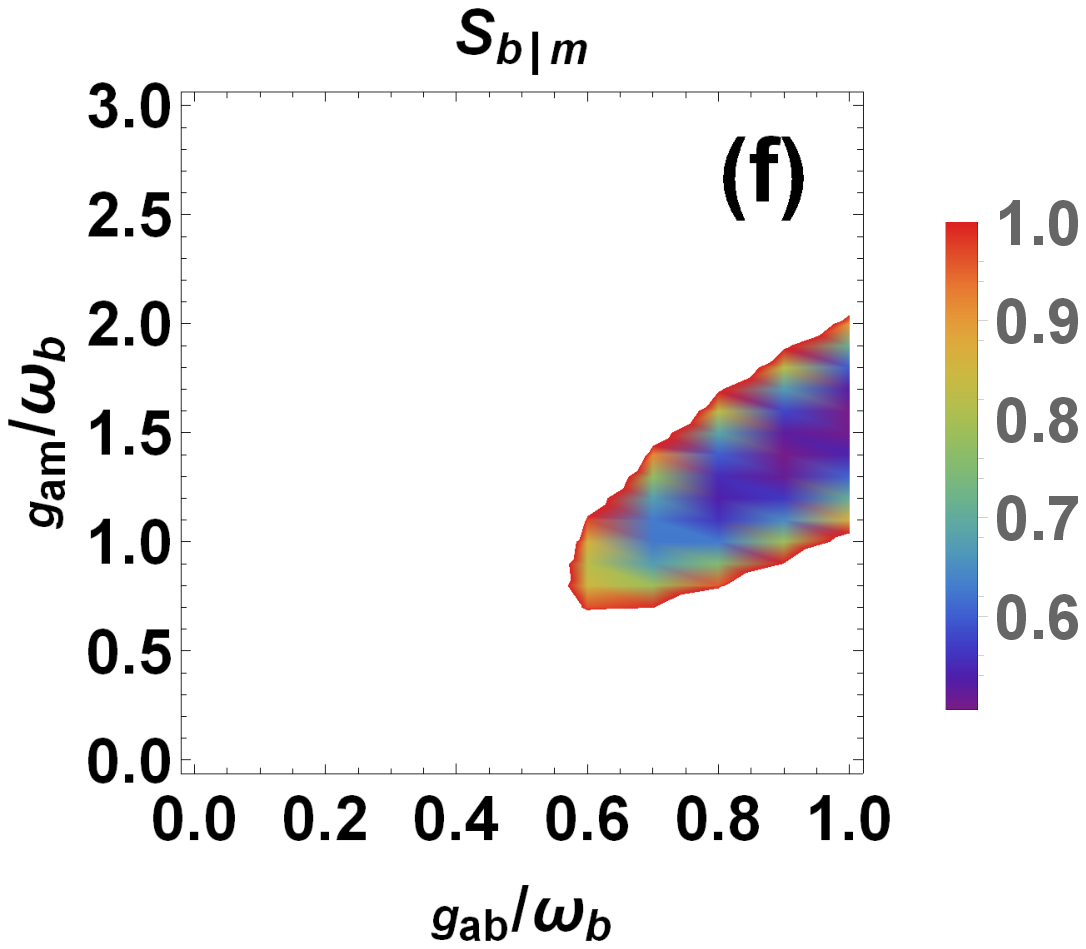}}}
\caption{Density plots of the entanglement $E_{bm}$ and steering $S_{m|b}$ and $S_{b|m}$ \emph{versus} the cavity dissipation rates $k_1$ and $k_2$ (a)-(c), and couplings $g_{am}$ and $g_{ab}$ (d)-(f). We take $g_{ab}=0.5g_{am}=\omega_m$ in (a)-(c), $\kappa_1=\kappa_2=10\omega_b$ in (d)-(f), $\Delta_m=-\omega_b$, $\Delta_1=-\Delta_2=\omega_b$, and the other parameters are the same as in Fig.\ref{f2}.}
\label{fig4}
\end{figure*}


\emph{\textbf{Phonon-magnon entanglement and steering.}}--The phonon-magnon entanglement can be witnessed when \cite{gio}
\begin{align}
E_{bm}=\frac{4V(\hat x_m^{\theta_m}+f_{x}\hat x_b^{\theta_b})V(\hat p_m^{\theta_m}-f_{y}\hat p_b^{\theta_b})}{(1+f_{x}f_{y})^2}<1,
\end{align}
where $V(\hat o)$ denotes the variance of the operator $\hat o$, and the angles $\theta_{b,m}$ and $f_{x,y}$ are used to minimize the variances, with $f_xf_y>0$. A tighter criterion is \cite{reid1}
\begin{align}
S_{m|b}=4V_{\rm inf}(\hat x_m^{\theta_m})V_{\rm inf}(\hat p_m^{\theta_m})<1,
\label{st1}
\end{align}
where $V_{\rm inf}(\hat x_m^{\theta_m})\equiv V(\hat x_m^{\theta_m}+f_{x} \hat x_b^{\theta_b})$ and $V_{\rm inf}(\hat p_m^{\theta_m})\equiv V(\hat p_m^{\theta_m}-f_{y} \hat p_b^{\theta_b})$ represent the inferred variances of the magnon mode, conditioned on the measurements of the mechanical position and momentum, with the optimal gains $f_{o}=V(\hat o_m^{\theta_m})-\langle \hat o_m^{\theta_m}\hat o_b^{\theta_b}\rangle/V(\hat o_b^{\theta_b})~(o=x,y)$. Eq.(\ref{st1}) shows that the Heisenberg uncertainty is seemingly violated, embodying the original EPR paradox \cite{bell,EPR}. Moreover, the conditional magnon squeezed states can be generated when Eq.(\ref{st1}) is hold, and it therefore reflects that the magnonic states can be steered by mechanics via the EPR entanglement and local measurements, characterizing quantum steering \cite{reid}, a type of quantum nonlocality \cite{wiseman} and originally termed by Sch\"{o}dinger in response to the EPR paradox \cite{sch}. Similarly, the reverse steering from the magnon to the phonon exists if
\begin{align}
S_{b|m}=4V_{\rm inf}(\hat X_b^{\theta_b})V_{\rm inf}(\hat Y_b^{\theta_b})<1.
\label{st2}
\end{align}
One-way steering is present when either of Eqs.(\ref{st1}) and (\ref{st2}) is hold. One-way property of quantum steering makes it intrinsically distinct from entanglement  and it is useful for, e.g., one-sided device-independent quantum cryptography \cite{qucry1}. Note that the smaller values of $E_{bm}$, $S_{m|b}$, and $S_{b|m}$ mean stronger entanglement and steering.

\begin{figure}[t]
\centerline{\scalebox{0.32}{\includegraphics{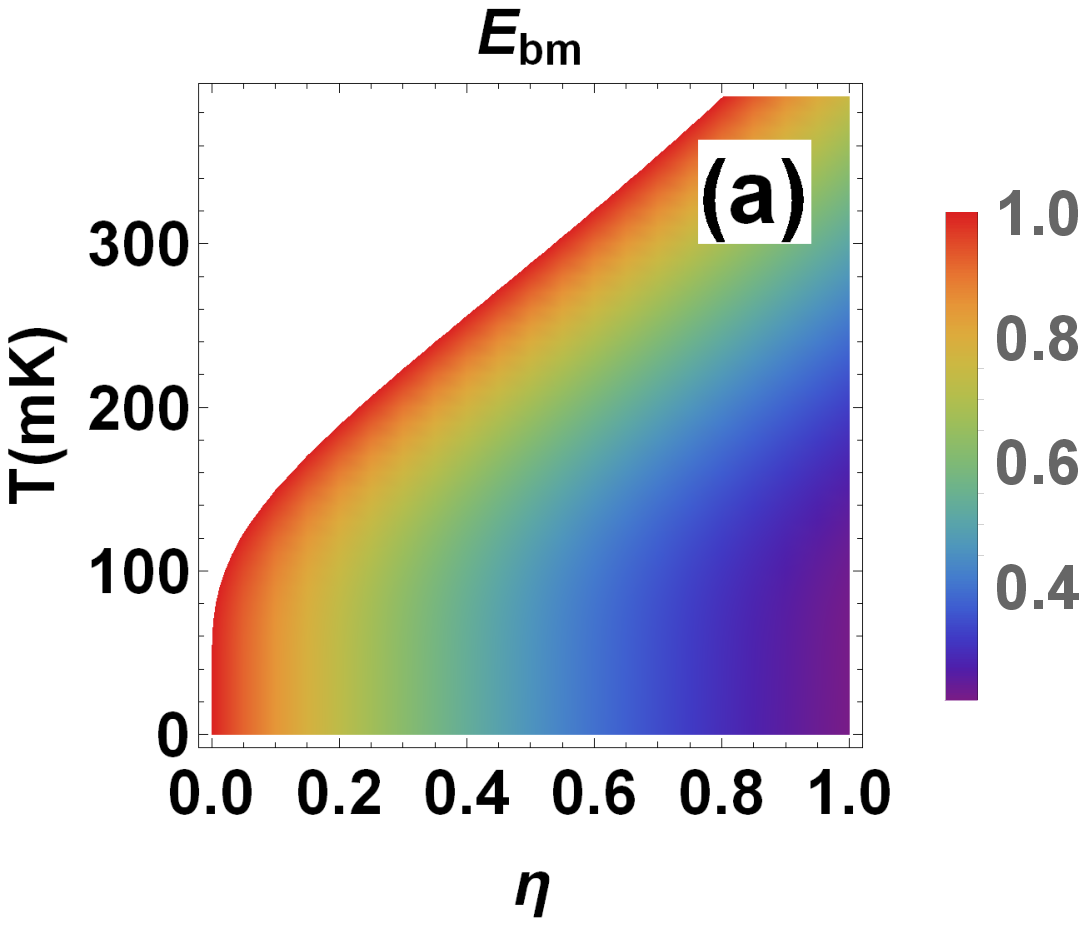}~~~~~~~~~~~~~\includegraphics{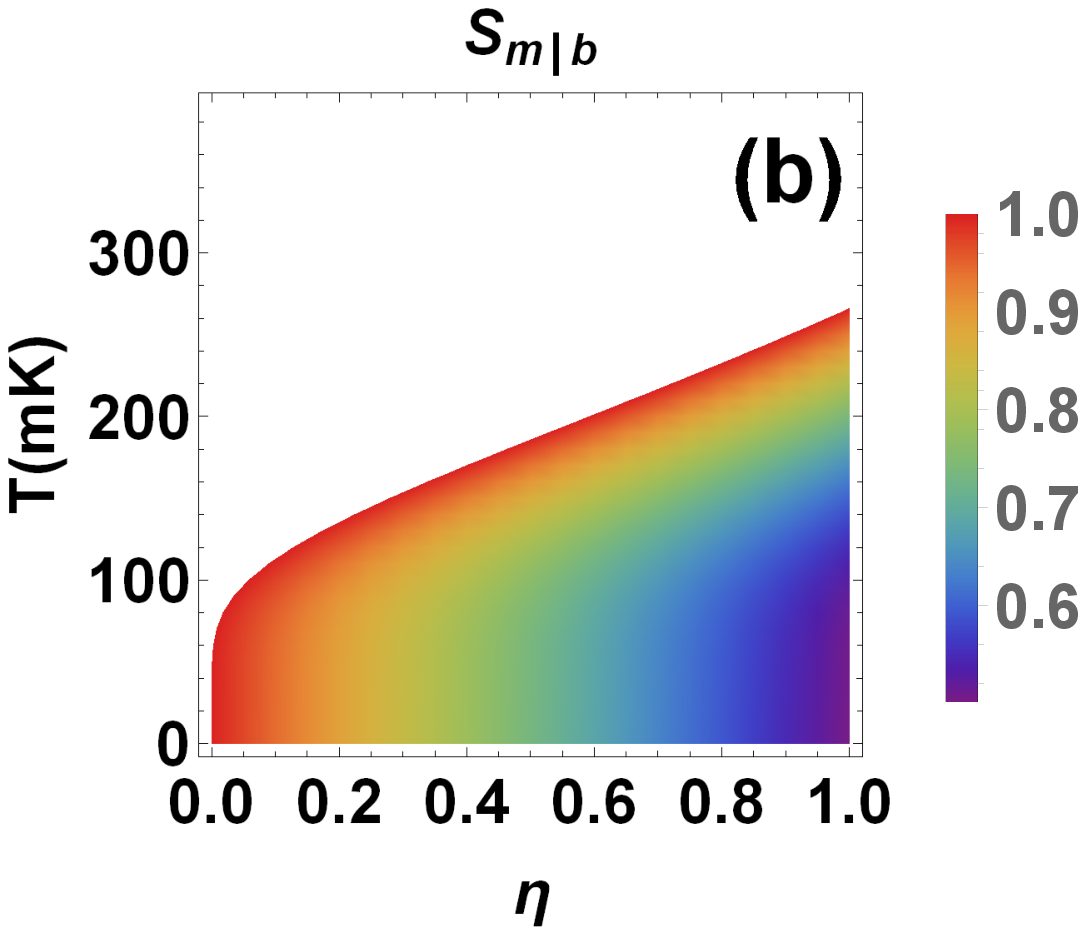}}}
\caption{Density plots of the entanglement $E_{bm}$ and steering $S_{m|b}$ \emph{versus} temperature $T$  and the cavity coupling efficiency $\eta$, with $\kappa_1=\kappa_2=10\omega_b$,
$\Delta_1=-\Delta_2=-\Delta_m=\omega_b$, $g_{ab}=0.5g_{am}=0.5\omega_b$, and the other parameters are the same as in Fig.\ref{f2}. The steering $S_{b|m}$ is absent and not plotted.}
 \label{fig3}
\end{figure}

\emph{\textbf{Results.}}--In Figs.\ref{f2}-\ref{fig4}, the dependence of the steady-state entanglement $E_{bm}$ and steering $S_{m|b}$ and $S_{b|m}$ on some key parameters are plotted. We adopt experimentally feasible parameters $\omega_b/2\pi=10$ MHz, $\omega_m/2\pi=10$ GHz, $\gamma_b/2\pi= 100~$ Hz, $\gamma_m/2\pi= 1.5$ MHz, and $T=T_{1,2}=T_{m,b}=30$~mK \cite{mpo2, uts, bis}.
We take the detuning $\Delta_1=\omega_b$ for cooling the mechanical mode, except for Figs.\ref{f2} (b) and (d).
We see that the EPR entanglement and steerings between the mechanical oscillator and the YIG sphere can be achieved in the steady-state regime. The phonon-to-magnon steering $S_{m|b}$ shows similar properties to the entanglement. By contrast, the reverse steering $S_{b|m}$ from the magnons to the phonons is absent, mainly due to a much larger magnon damping rate than that of the mechanics, i.e., $\gamma_m\gg\gamma_b$, and it is merely present for unbalanced cavity dissipation rates $\kappa_1$ and $\kappa_2$ or relatively large couplings $g_{ab}$ and $g_{am}$, as shown in Fig.\ref{fig4} (c) and (f). Moreover, the steering $S_{m|b}$ is stronger than the reverse $S_{b|m}$ when both of them are present.

Specifically, as shown in Fig.\ref{f2} the entanglement and steering
become maximal under strong cavity dissipation, $\kappa_{1,2}\gg \{g_{ab}, g_{am}, \omega_b\}$,
far beyond the sideband-resolved regime of the electromechanical system,
and they are also optimized when the detuning $\Delta_m=-\omega_b$.
This can be understood as follows: the phonon-magnon entanglement in fact originates from the photon-phonon entanglement built up by the electromechanical coupling. As studied by one of us in Ref.\cite{tan}, on the bad-cavity condition $\kappa_1\gg\omega_b$, the stationary entanglement between the mechanical oscillator and the cavity output photons at frequency $\omega_d-\omega_b$ becomes maximal and much stronger than the intracavity-photon-phonon entanglement which does not yet exhibit quantum steering; via the unidirectional photon-photon coupling and photon-magnon coupling, the output photon-phonon entanglement is then partially transferred into the phonon-magnon entanglement. Since the transfer efficiency depends on the product $\kappa_1\kappa_2$ and resonates also at frequency  $\omega_d-\omega_b$ for the photons and magnons in the second cavity, the phonon-magnon entanglement is therefore maximal for the bad cavities and at the detuning $\Delta_m=-\omega_b$.

Figure 2 also shows that under the bad-cavity condition, the steady-state entanglement and steering are attainable in relatively wide ranges of the detunings $\Delta_{1,2}$. They are present in the red-detuned regime of the electromechanical subsystem ($\Delta_1>0$) for the stability consideration and become maximum at the instability threshold $\Delta_1\approx0$. We also see that exactly due to the strong cavity coupling $\sqrt{\kappa_1\kappa_2}$, the entanglement and steering can still be achieved even for largely different two cavity frequencies (i.e., $\Delta_1\neq\Delta_2$), demonstrating their robustness against the frequency dismatch between the two cavities.
In addition, as depicted in Fig.\ref{fig4} the optimal entanglement and steering are obtained for the cooperativity parameters $\mathcal C_b=g_{ab}^2/\kappa_1\gamma_b\approx 6\times10^3$ and $\mathcal C_m=g_{am}^2/\kappa_2\gamma_m\approx 1$, which can readily be realized with current state-of-the-art experimental technology \cite{uts, bis}.

Fig.\ref{fig3} reveals that the entanglement and steering are robust against the inefficiency of the unidirectional coupling and thermal fluctuations. They can survive up to $T>100~m$K for a realistic coupling efficiency $\eta=0.5$, and the entanglement is more robust than the steering, since the latter embodies a stronger quantum correlation. We note that in the current experiments in microwave domain \cite{sid,jod}, the cascade-cavity-coupling efficiency up to $\eta\approx0.75$ can be achieved. For the YIG sphere, the cooling to $10$~mK $\sim$ $1$~K by using a dilution refrigerator, merely with small line broadening, has been achieved in the experiment \cite{mpo2}. In addition, around the temperature $T_{1,2}=T_b=T_m=30$~mK by using cryostat in the experiments \cite{mpo2, bis, uts}, the mean thermal magnon number $\bar n^{\rm th}_{m}\approx 0$, while the mean thermal phonon number $\bar n^{\rm th}_b\approx 60$.

\emph{\textbf{Discussion and Conclusion.}}-- For the parameters considered $\omega_b/2\pi \,\,{=}\,\,10$~MHz, $\omega_m/2\pi \,\,{=}\,\, 10$~GHz, $\Delta_1 = {-} \Delta_2 = {-} \Delta_m = \omega_b$, and $\kappa_{1}=10\omega_b$, to achieve the electromechanical coupling $g_{ab}\equiv \tilde{g}_{ab}\sqrt{\mathcal P_d\kappa_1}/\sqrt{\hbar \omega_{c_1}(\Delta_1^2+\kappa_1^2/4)}=0.5\omega_b$, the pumping power $\mathcal P_d=1~\mu$W is required, given the single-photon electromechanical coupling $\tilde {g}_{ab}/2\pi\approx 150$ Hz \cite{uts}. When a 400-$\mu$m-diameter YIG sphere is considered, the number of spins in the sphere $N\approx 7\times 10^{16}$, with the net spin density of the sphere $\rho_s=2.1\times 10^{27}~\text m^{-3}$, and the electromagnonical coupling $g_{am}\equiv g_{m0}\sqrt{N}=\omega_b$ can be obtained for the single-spin coupling $g_{m0}/2\pi\approx 38$~mHz \cite{mpo2}. In addition, for the given couplings $g_{ab}$ and $g_{am}$, the mean magnon number $|\langle \hat m\rangle|^2\approx3.4\times 10^8\ll 2N s$ for the spin number $s=\frac{5}{2}$ of the ground-state ferrum ion $\text {Fe}^{3+}$, which ensures the condition of low-lying excitation of the spin ensemble necessary for the present model.

As for the detection of the EPR entanglement and steering, we adopt the well established method \cite{vitali} by coupling the mechanical and magnon modes to the separate probing fields $\hat p_b$ and $\hat p_m$ (see Fig.1). When the probing cavity $\hat p_b$ is driven by a {\it weak} red-detuned pulse and the cavity $\hat p_m$ is resonant with the magnon mode, the beam-splitter-like Hamiltonian $\hat H_{bp}=g_{bp}(\hat b\hat p_b^\dag+\hat b^\dag\hat p_b)$ and $\hat H_{mp}=g_{mp}(\hat m\hat p_m^\dag+\hat m^\dag\hat p_m)$ are thus activated. 
Therefore, the states of the mechanical and magnon modes can be transferred onto the probes. By homodyning the outputs of the probes and measuring the variances of the quadratures, one can verify the entanglement and steering.

In conclusion, we present a deterministic scheme for creating hybrid EPR entanglement channel between a macroscopic mechanical oscillator and a magnon mode in a distant macroscopic YIG sphere. 
The entanglement is created in the electromechanical cavity and distributed remotely to the electromagnonical cavity by coupling the output field of the former to the latter. Strong stationary phonon-magnon EPR entanglement and steering can be achieved under realistic parameters far beyond the sideband-resolved regime. These features clearly distinguish from the existing proposals \cite{li1,li3} where the entanglement is generated locally under the condition of resolved sidebands, and thus the present proposal manifests its unique advantages. With the rapid progress in realizing the coupling between hybrid massive systems \cite{plzk2, ham}, our proposal is promising to be realized in the near future. This distant hybrid EPR entanglement and steering between two truly massive objects may find its applications in the fundamental study of macroscopic quantum phenomena, as well as in quantum networking and one-sided device-independent quantum cryptography. 
Further investigations would conclude quantum-state exchange between a mechanical oscillator and a distant YIG sphere.



\section*{Acknowledgment}
This work is supported by the National Natural Science Foundation of China (No.11674120), the Fundamental Research Funds for the Central Universities (No. CCNU18TS033), and the European Research Council Project (Grant No. ERC StG Strong-Q, 676842).



\end{document}